\documentclass{sf2a-conf2018}
\usepackage{graphicx}
\usepackage{hyperref}
\usepackage[]{natbib}  
\usepackage{epstopdf}

\def\BibTeX{{\rm B\kern-.05em{\sc i\kern-.025em b}\kern-.08em
    T\kern-.1667em\lower.7ex\hbox{E}\kern-.125emX}}
\bibpunct{(}{)}{;}{a}{}{,}  

\begin{document}

\TitreGlobal{SF2A 2018}


\title{Analysis of HST, VLT and Gemini coordinated observations of Uranus late 2017 : a multi-spectral search for auroral signatures}

\runningtitle{Analysis of recent Uranus UV/IR observations}

\author{L. Lamy}\address{LESIA, Observatoire de Paris, PSL, CNRS, 5 Place Jules Janssen, 92195 Meudon, France.}
\author{C. Berland$^1$}
\author{N. Andr\'e}\address{IRAP, CNRS, UniversitŽ Paul Sabatier, 9 Avenue du Colonel Roche, 31400 Toulouse, France.}
\author{R. Prang\'e$^1$}
\author{T. Fouchet$^1$}
\author{T. Encrenaz$^1$}
\author{E. Gendron$^1$}
\author{X. Haubois}\address{ESO, Alonso de Cordova 3107, Vitacura, Casilla 19001 Santiago de Chile, Chile.}
\author{C. Tao}\address{National Institute of Information and Communications Technology, Tokyo, Japan.}
\author{T. Kim}\address{CSPAR, University of Alabama in Huntsville, USA.}




\setcounter{page}{237}


\maketitle


\begin{abstract}
On 6 Sept. 2017, an exceptional coronal mass ejection departed from the Sun toward the Earth and Uranus, whose magnetospheres are sensitive to the solar wind. The resulting interplanetary shock triggered geomagnetic storm and intense aurora at Earth the next day and was predicted by MHD models to reach Uranus around 10-11 Nov. This event provided a unique opportunity to investigate the auroral response of the asymmetric Uranian magnetosphere in its intermediate equinox-to-solstice configuration. Coordinated multi-spectral observations were acquired with the Hubble Space Telescope (HST) in the far-UV (FUV), with the Very Large Telescope (VLT) and Gemini North in the near-IR (NIR) and with Chandra in the X-ray domain. In this study, we focus on the analysis of NIR images obtained between 9 and 17 Nov. 2017 which are compared to one FUV image acquired on 11 Nov. The latter reveals a bright southern auroral spot in the H$_2$ bands, which we use as a reference to locate auroral precipitations. The NIR images were aimed at mapping H$_3^+$ emission from the Uranian ionosphere and at updating the results built from a couple of pioneer images taken 25 years ago. These new high resolution images reveal H$_3^+$ from the whole disc although brighter near the southern pole, but show no evidence of localized auroral emission. 
\end{abstract}



\section{Introduction}
Uranus hosts a large-scale magnetic field, whose interaction with the solar wind (SW) generates a magnetosphere, together with auroral emissions at radio and ultraviolet (UV) wavelengths which were detected in 1986 at solstice during the flyby of the planet by the Voyager 2 (V2) spacecraft \citep{Ness_Science_86,Warwick_Science_86,Broadfoot_Science_86}. The V2 UV Spectrometer revealed patchy aurorae at H-Lyman $\alpha$ and in the H$_2$ bands collisionally excited in the upper atmosphere by the precipitation of energetic electrons near the magnetic poles. More than two decades later, the Uranian aurorae were re-detected in the Far-UV (FUV) by the Space Telescope Imaging Spectrograph (STIS) of the Hubble Space Telescope (HST) using a novel approach \citep{Lamy_GRL_12,Barthelemy_Icarus_14}. These observations were conducted past equinox in 2011, 2012, 2014 during active SW conditions at Uranus, predicted in advance by magneto-hydrodynamic (MHD) models \citep{Lamy_JGR_17}. The STIS observations revealed transient spots around both magnetic poles radiating a few kR of photons (a few GW of power) in the H$_2$ bands, and attributed to a time-variable magnetosphere/SW interaction. 


Another key element of the Uranian system is the coupling of the magnetosphere with the ionosphere at the footprint of auroral field lines. At Uranus, such a coupling is complex due to the large tilt between the ionosphere dragged by the planetary rotation and the magnetosphere rotating around a highly tilted magnetic axis. Signatures of this coupling can be searched for in near-IR (NIR) ro-vibrational emissions of the ionospheric, thermally excited, H$_3^+$ ion, whose density is increased by auroral particle precipitations. However, V2 did not carry any NIR instrument. Long-term NIR spectroscopic observations of Uranus performed from Earth from the 1990s on revealed roughly uniform H$_3^+$ emission from the disc, mainly attributed to Extreme-UV (EUV) solar-driven ionization of the upper atmosphere in a $\sim$600-800~K hot thermosphere. The H$_3^+$ total emission, column density and temperature additionally displayed a long-term variability roughly consistent with the solar cycle, together with a few significant excursions tentatively attributed to transient auroral precipitations \citep{Trafton_ApJ_93,Trafton_ApJ_99,Melin_ApJ_11,Melin_Icarus_13}. A limited set of NIR images with narrow-banded filters tracking H$_3^+$ lines was also taken in 1993 by the IRTF 3-m telescope and analyzed by \citet{Lam_ApJ_97}. Despite a low signal-to-noise ratio, these images interestingly revealed an intriguing H$_3^+$ northern bright region seemingly corotating across the disk. The authors could not conclude about its auroral nature, while setting up an upper limit of $20\%$ for the contribution of any auroral process to the total H$_3^+$ signal, a few 10$^{-26}$~W.m$^{-2}$ once disc-averaged. 

On 6 Sept. 2017, a powerful coronal mass ejection departed from the Sun toward the Earth and Uranus. This fast propagating interplanetary shock triggered geomagnetic storm and intense aurora at Earth on 7 Sept. and was predicted by two MHD propagation codes to reach Uranus around 10-11 Nov. Coordinated observations of Uranus, allocated on director's time, were acquired with the Hubble Space Telescope (HST) in the far-UV (FUV), with the Very Large Telescope (VLT) and Gemini North in the near-IR (NIR) and with Chandra in the X-ray domain to track any auroral response of the asymmetric magnetosphere in its new equinox-to-solstice configuration. In this study, we focus on the analysis of six NIR observing sequences and one FUV image obtained between 9 and 17 Nov. 2017. The exhaustive analysis of the full set of FUV observations, together with that of X-ray data, is beyond the scope of this paper.

\begin{figure}[ht!]
\centering
\includegraphics[width=1.0\textwidth,clip]{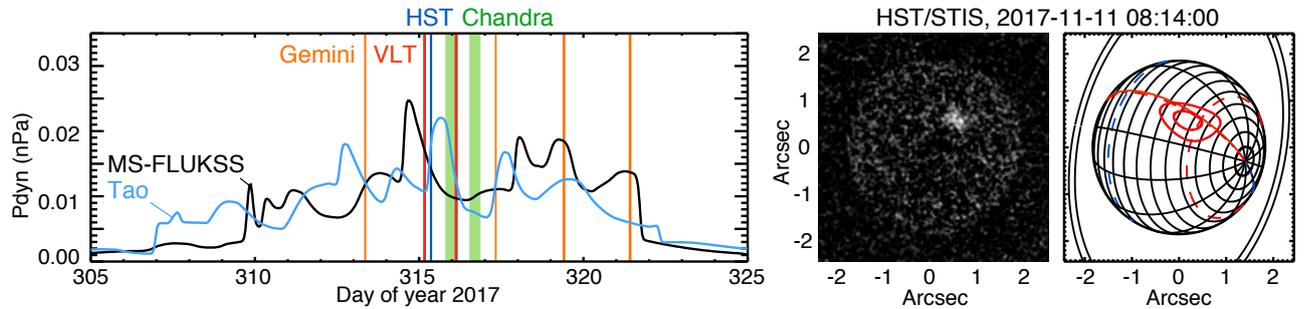}      
\caption{{\bf Left:} Solar wind dynamic pressure at Uranus from 1 Nov. to 21 Nov. 2017 (day of year 305 to 325). The vertical colored lines mark the timing of coordinated observations. {\bf Right:} HST/STIS background-subtracted image and planetary configuration on 11 Nov. 2017. The model southern auroral oval (red curves) fits the auroral spot.}
\label{fig1}
\end{figure}


\section{Solar wind conditions at Uranus in November 2017}

The method employed to predict the SW conditions at Uranus is described at length in \cite{Lamy_JGR_17}. We just remind that the SW parameters, measured at Earth, are propagated up to Uranus with two distinct MHD models : the MS-FLUKKS model \citep{Kim_ApJ_16} and the Tao model \citep{Tao_JGR_05}, with a typical uncertainty of $\pm2$ days. The modeled interplanetary dynamic pressure at Uranus is plotted on Figure \ref{fig1} (left) from 1 Nov. to 21 Nov. 2017. This interval witnesses a large-scale jump in dynamic pressure rising from $\sim0.001$~nPa on 2 Nov. up to $0.025$~nPa on 10 Nov. (MS-FLUKKS, black line) or $0.022$~nPa on 11 Nov. (Tao model, blue line), followed by a recovery phase lasting approximately one week. The amplitude of this predicted pressure front stands as the largest ever sampled at Uranus to date (see Fig. 1 of \cite{Lamy_JGR_17}). 

\section{HST detection of FUV southern aurorae}
A series of FUV, NIR (and X-ray) imaging observations were executed from 9 to 17 Nov. to sample any auroral response of the Uranian magnetosphere to this strong SW shock interaction. They are labelled with vertical lines in Fig. \ref{fig1} (left). HST/STIS imaged Uranus once during this interval, at 08:14:00~UT on 11 Nov., near the pressure peak. Figure \ref{fig1} (right) displays the background-subtracted STIS image, derived from a 2563~s-long exposure obtained with the clear filter, and the associated planetary configuration. The image clearly reveals a bright southern auroral spot, reminiscent of those previously observed past equinox, although somewhat brighter and more spatially extended. We fitted this auroral feature with a southern model auroral oval, as defined in \citep{Lamy_JGR_17}, to reference the locus of auroral regions rotating in both hemispheres over the full interval.

\begin{figure}[ht!]
 \centering
 \includegraphics[width=1.0\textwidth,clip]{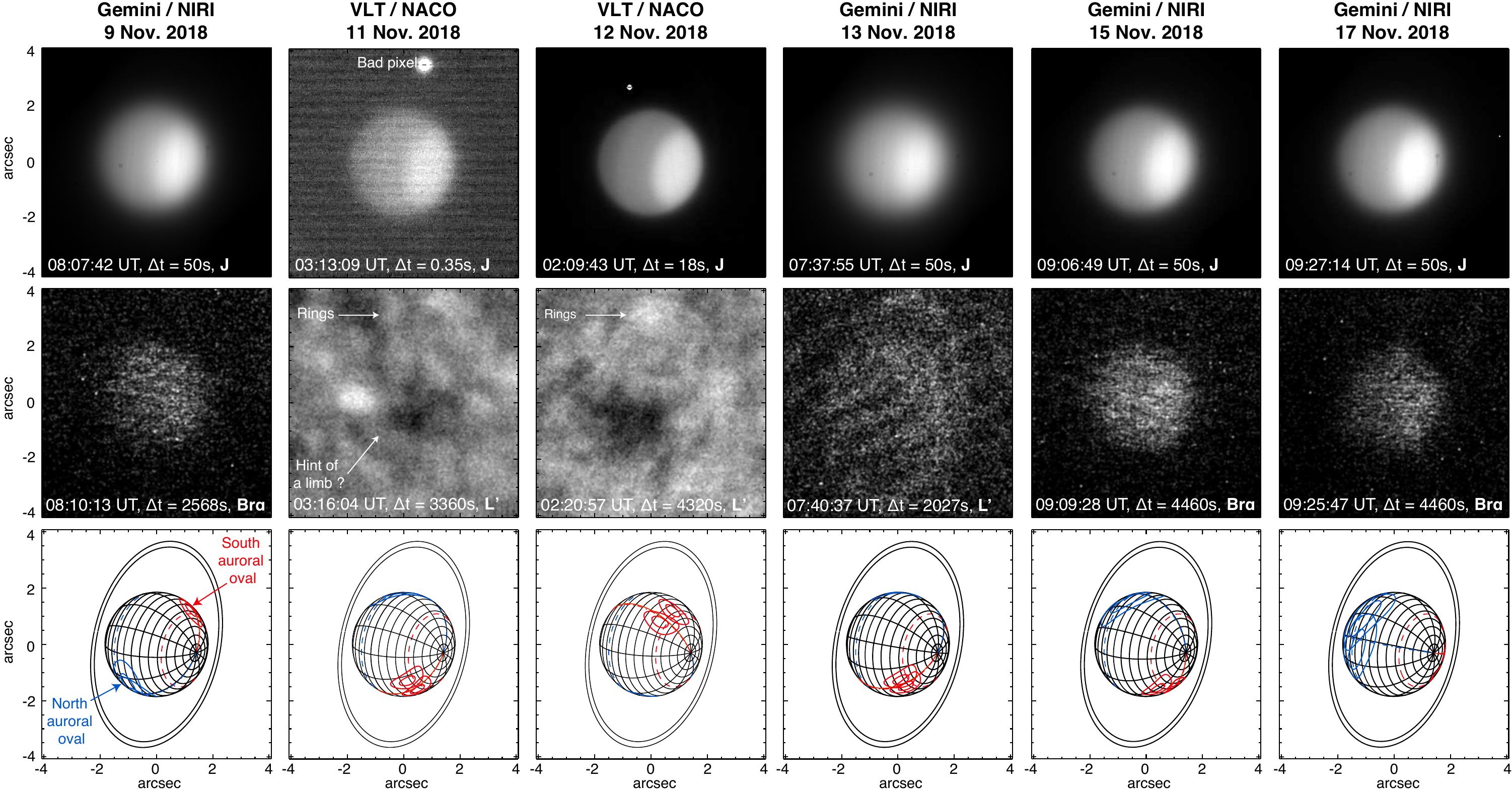}      
\caption{NIR images of Uranus obtained with VLT/NACO and Gemini/NIRI from 9 to 17 Nov. 2017. \textbf{Top:} Acquisition images, taken in the J band ($1.15-1.33~\mu$m for NACO, $1.14-1.39~\mu$m for NIRI). \textbf{Middle:} Science images, taken with the wideband L' filters ($3.49-4.11~\mu$m for NACO, $3.43-4.13~\mu$m for NIRI) and the narrowband Br$\alpha$ one for NIRI only ($3.95-4.02\mu$m). \textbf{Bottom:} Grids of planetocentric coordinates at the limb. The red/blue dashed lines indicate the latitude of southern/northern magnetic poles. The red/blue solid curves indicate southern/northern model auroral ovals (each plotted twice, at the beginning/end of the NIR exposure).}
\label{fig2}
\end{figure}

\section{NIR images and map of $H_3^+$ emission}

VLT and Gemini North observed Uranus during two and four nights, respectively, with different instrumental configurations. The VLT images were acquired from Paranal (Chile) with the NaCo Nasmyth Adaptive Optics System Near-Infrared Imager and Spectrograph (NACO\footnote{https://www.eso.org/sci/facilities/paranal/instruments/naco.html}) instrument available on the Antu 8.2-m unit telescope. Using adaptative optics, VLT/NACO acquired short exposures with the J filter ($1.15-1.33~\mu$m) for acquisition images and long exposures with the L' broadband filter ($3.49-4.11~\mu$m) encompassing many H$_3^+$ lines. The Gemini images were acquired from Mauna Kea in Hawai (USA) with the Near-InfraRed Imager and Spectrograph (NIRI\footnote{https://www.gemini.edu/sciops/instruments/niri/}) instrument of the Gemini North 8.1-m telescope. Without using adaptative optics, Gemini/NIRI acquired short exposures with the J filter ($1.14-1.39~\mu$m) for acquisition images and long exposures with two filters for science images : the broadband L' filter ($1.14-1.39~\mu$m) and the narrowband Br$\alpha$ one ($3.95-4.02\mu$m, only available on NIRI), matching a limited number of H$_3^+$ lines.

The processed, background-subtracted images are displayed on Figure \ref{fig2}. The acquisition images, best illustrated by the VLT image of 12 Nov., clearly display solar reflected emission from the disc much brighter around the southern pole, typically beyond $\sim-40^\circ$ latitude. The VLT/NACO L' images of 11 and 12 Nov. show evidence of solar reflected emission from the rings and emission from the disc, which are obvious in the Gemini/NIRI L' image of 13 Nov. Despite its lower spatial resolution, Gemini/NIRI was likely more sensitive owing to a lower sky background temperature (about 0$^\circ$C in Mauna Kea vs $\sim25^\circ$C in Paranal). According to the low planetary albedo of Uranus at the sampled NIR wavelengths \citep{Lam_ApJ_97}, the signal from the disc is attributed to H$_3^+$ emission.

\begin{figure}[ht!]
 \centering
 \includegraphics[width=0.45\textwidth,clip]{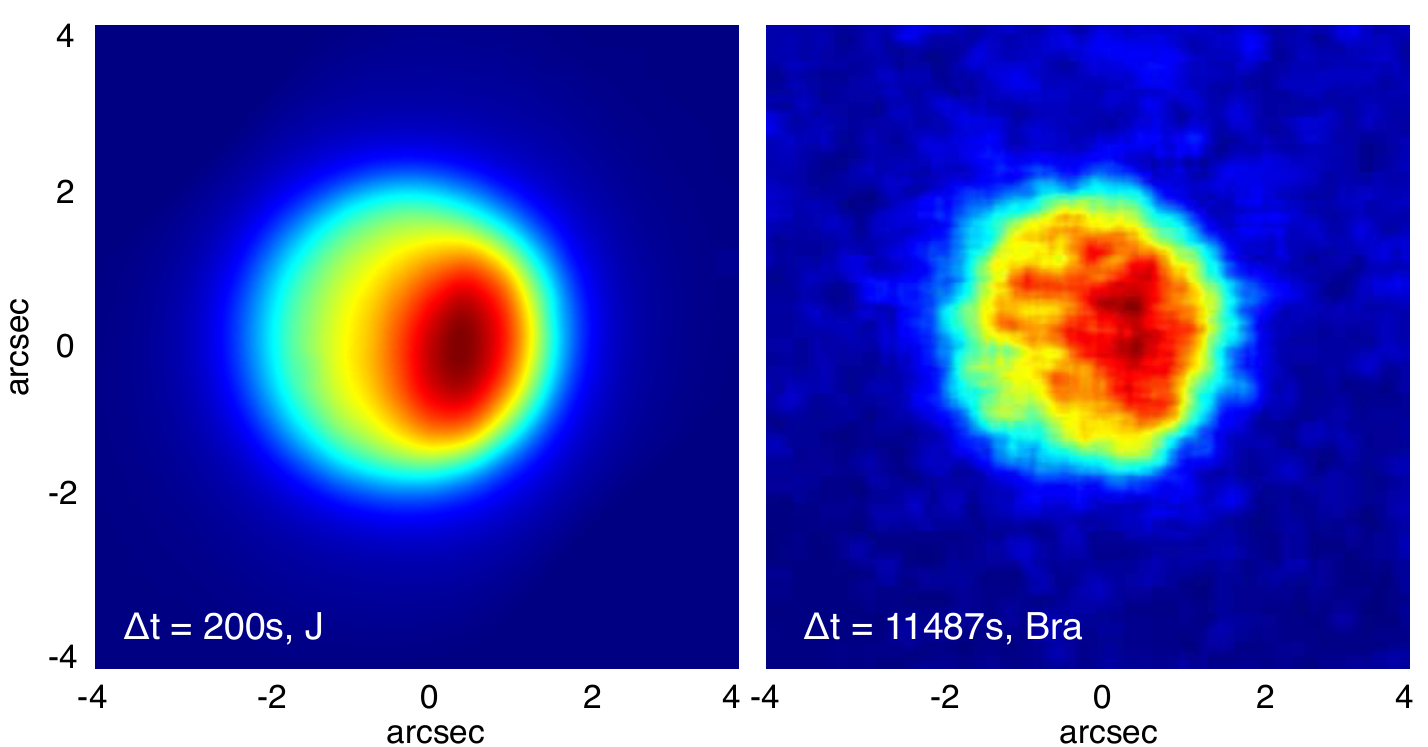}      
\caption{\textbf{Left:} Average Gemini/NIRI J image, smoothed for comparison. \textbf{Right:} Average Gemini/NIRI Br$\alpha$ image. Both images display a similar spatial distribution of intensity, with a local maximum around the southern pole.}
\label{fig3}
\end{figure}

The Gemini/NIRI Br$\alpha$ images taken on 9, 15 and 17 Nov. clearly confirm and precisely map the H$_3^+$ atmospheric emission, while illustrating the gain in sensitivity with narrowbanded filters. The images reveal that H$_3^+$ is non-uniformly radiating across the disc, with brighter emission around to the south pole. They additionally show the absence of localized emissions associated with model auroral ovals derived from Figure \ref{fig1} (right). The low signal-to-noise ratio prevented us from investigating any temporal variability. A more comprehensive picture is brought by Figure \ref{fig3}, which displays average Gemini/NIRI images for the J and Br$\alpha$ filters, corresponding to 200~s and 11487~s effective integration time. The average J image was smoothed to facilitate cross-comparison with the L' one. The spatial correspondance between both images, with a similar inhomogeneous spatial distribution and suggests that H$_3^+$ emission is linked to atmospheric processes governing the albedo. 


\section{Discussion and perspectives}

These new NIR images of Uranus reveal two important informations. First, they do not reveal any transient localized aurora while one single FUV image sufficed to do so over the same interval. Second, they confirm that H$_3^+$ is non-uniformly radiating across the disc, with brighter emission near the southern pole. This somewhat differs from the results obtained by \cite{Lam_ApJ_97} 25 years ago. This H$_3^+$ distribution likely diagnoses a hotter thermosphere around the southern pole when approaching summer. Interestingly, it also compares to the distribution of the solar reflected emission which precisely brightens beyond $-40^\circ$ latitude. In a parallel study, \cite{Toledo_GRL_18} linked this bright polar cap to strong methane absorption rather than to any aerosol abundance such as any stratospheric haze arising from auroral precipitations. Overall, the analysis of NIR spectroscopic observations of Uranus should take into account the non-uniform H$_3^+$ emission.








\begin{acknowledgements}
The authors thank the HST, VLT and Gemini directors for the allocated observations and support from CNES and CNRS/INSU. The analyzed data were obtained from the HST program GO/DD 15380, the VLT program 2100.C-5006(C) and the Gemini program GN-2017B-DD-6. LL thanks R. Galicher and A. Boccaletti for fruitful discussions about the NIR data processing and H. Kita, Y. Kasaba and T. Sakonoi for attempting to get support observations from IRTF. Support for the HST program was provided through a NASA/STSci grant, operated by AURA under NASA contract NAS5-26555.
\end{acknowledgements}

\end{document}